# Two-fold symmetric superconductivity in the kagome superconductor RbV$_3$Sb$_5$

Shuo Wang[1,2,3,†], Jing-Zhi Fang[1,2,†], Ze-Nan Wu[1,2,†], Sirong Lu[1,2,3], Zhongming Wei[4,5], Zhiwei Wang[6,7], Wen Huang[1,2,3], Yugui Yao[6,7], Jia-Jie Yang[1,2], Tingyong Chen[1,2,3], Ben-Chuan Lin[1,2,3,*], Dapeng Yu[1,2,3]

[1]*Shenzhen Institute for Quantum Science and Engineering, Southern University of Science and Technology, Shenzhen, 518055, China*

[2]*International Quantum Academy, Shenzhen 518048, China.*

[3]*Guangdong Provincial Key Laboratory of Quantum Science and Engineering, Southern University of Science and Technology, Shenzhen, 518055, China*

[4]*State Key Laboratory of Superlattices and Microstructures, Institute of Semiconductors, Chinese Academy of Sciences, Beijing 100083, China*

[5]*Center of Materials Science and Optoelectronics Engineering, University of Chinese Academy of Sciences, Beijing 100049, China*

[6]*Key Laboratory of Advanced Optoelectronic Quantum Architecture and Measurement, Ministry of Education, School of Physics, Beijing Institute of Technology, Beijing 100081, China*

[7]*Micronano Center, Beijing Key Lab of Nanophotonics and Ultrafine Optoelectronic Systems, Beijing Institute of Technology, Beijing 100081, China*

[†] *These authors contributed equally to this work.*

[*]linbc@sustech.edu.cn

**Abstract**

The recently discovered kagome superconductors offer a promising platform for investigating intertwined orders and novel states, including topology, superconductivity, charge density waves, and more. The interplay among these orders can spontaneously break rotational symmetry, giving rise to exotic phenomena such as nematicity or even nematic superconductivity. Here we present our findings on the two-fold symmetric superconductivity in thin-flake RbV$_3$Sb$_5$ in response to direction-dependent in-plane magnetic fields, in contrast to the inherent six-fold structural symmetry of the crystal lattice. The two-fold symmetry was evidenced through a combination of magnetoresistance transport experiments, critical magnetic field measurements, and observations of anisotropic superconducting gaps. Additionally, by altering the experimental configuration, we also detected the presence of six-fold symmetric components superimposed on the two-fold symmetry at the boundary between normal and superconducting states. Our results underscore the correlation-driven symmetry-breaking phenomena and emphasize the potential of this correlated kagome family as a promising platform for investigating intertwined orders, including unconventional superconductivity.

# Introduction

The recent discovery of kagome superconductors $AV_3Sb_5$ (A=K, Rb, Cs) revives the interest in unconventional superconductivity and other correlation-driven electronics states in quantum materials[1-7]. The distinct kagome lattice structure offers a rich landscape of exotic phenomena, including spin liquids, charge or spin density waves, non-reciprocal charge transport[8], novel superconductivity[9-12], and topological phases[13]. Notably, in this system, the time-reversal symmetry was reported to be possibly spontaneously broken[4,7], sparking optimism for the emergence of unconventional superconductivity below the superconducting transition temperature[3,7,14-17]. A robust zero bias conductance bias peak inside the vortex core of $CsV_3Sb_5$[15], the superconducting dome[16,18], possible spin-triplet superconductivity[19], and the roton pair density wave[17] of kagome superconductors were further reported, implying the possible unconventional superconducting pairing mechanism and intertwined orders in the kagome superconductors[7,9-12]. In addition, the charge density wave (CDW) driven nematicity was observed below the critical temperature of 35 K[6]. Thus, it opens up the fascinating prospect of studying the interplay between nematicity and superconductivity, since nematic superconductivity can act as a probe of unconventional superconducting pairing[20-25] or even topological superconducting states[26]. However, up to now, few experiments of directly measuring the in-plane nematic phase under the critical superconducting temperature have been reported within the kagome family of materials, especially in transport experiments. The nature and symmetry of the pairing state in kagome superconductors $AV_3Sb_5$[14,15,17-19,27] remains an area of ongoing investigation.

In this work, we report the two-fold symmetry of superconducting states in the thin-film $RbV_3Sb_5$ sample when subjected to direction-dependent, in-plane magnetic fields. Such periodicity is evidenced through a combination of magnetoresistance experiments, measurements of the effective critical field below the critical superconducting temperature, and the measurements of the critical current revealing the anisotropic superconducting gap[25]. The two-fold symmetry disappears when the sample enters into the normal state and possible trivial effects were carefully ruled out. Furthermore, the six-fold symmetry superimposed on the two-fold symmetry was observed with different configurations. This anisotropic superconducting gap, in contrast to the isotropic superconducting gap typically associated with s-wave superconductivity, indicates the possible unconventional superconductivity of the thin-film kagome superconductor $RbV_3Sb_5$ and further highlights the exotic orders in this family of kagome materials.

# Results

**Anisotropic superconducting behavior**

RbV$_3$Sb$_5$ is a layered kagome superconductor composed of alternately stacked V-Sb slabs and Rb layers[5]. Each V-Sb slab has a 2D kagome net of V atoms forming the hexagonal structure with centers occupied by Sb atoms as shown in Fig. 1a. Obviously, such a structure would indicate a C$_6$ rotational symmetry[6]. Fig. 1b shows the original thin-film RbV$_3$Sb$_5$ sample, which was then encapsulated with the hexagonal boron nitride (h-BN) in the glovebox filled with N$_2$ gas (< 0.01 ppm O$_2$ and H$_2$O concentrations) to prevent any possible contamination and preserve the pristine nature of the RbV$_3$Sb$_5$ device. The thickness of the various devices in our study ranges from 20 nm to 40 nm. The charge density wave (CDW) transition, indicated in Fig. S1, takes place at around 100 K[28]. The temperature dependence of the differential resistance is presented in Fig. 1c. As the temperature increases, the critical current, a measure of the superconducting gap, gradually decreases until superconductivity is completely suppressed at higher temperatures. In Figs. 1d-f, we depict the temperature dependence of the magneto-resistance R$_{14,23}$, where voltage measurements were taken between terminals 2 and 3, while the current was applied along terminals 1 and 4. This measurement reveals a noticeable transition from the normal state, characterized by finite resistance, to the superconducting state with zero resistance. The contour in the figures delineates the boundary between the superconducting and normal states, providing insight into the temperature dependence of the critical magnetic field.

It is apparent that the critical magnetic fields along different axes exhibit pronounced anisotropic behavior. In this context, the in-plane critical magnetic field is notably larger than the out-of-plane counterpart, a characteristic commonly observed in thin-film superconductors due to the limited diamagnetic energy within the thin film compared to bulk superconductors[29]. However, what sets our findings apart is that, unlike most superconductors, the in-plane critical magnetic fields measured along different directions exhibit significant disparities in their anisotropy. If we define the critical magnetic field Hc$_2$ corresponding to half the value of normal resistance, then the anisotropy can be quantified as Hc$_{2x}$/Hc$_{2y}$ ~ 3.56, representing a conspicuous anisotropic behavior[23–25]. In other words, the rotational symmetry is broken, as θ = 0° and 90° are no longer equivalent.

**Two-fold symmetric superconductivity**

Figs. 2a&b demonstrate the in-plane anisotropy of the resistance along different directions. A rapid decrease in critical current (almost exponential) was observed as the magnetic field increased perpendicular to the current direction (θ = 90°)[30]. This observation signifies that superconductivity is considerably more susceptible to suppression when the magnetic field is aligned with θ = 90°, compared to when it aligns with θ = 0°. This distinctive anisotropic behavior further suggests

anisotropy in the superconductor gap symmetry. In light of these findings, we conducted magneto-transport measurements with in-plane magnetic fields rotated from 0° to 360°, with 0° denoting the direction of the magnetic field aligned with the current. The results, as illustrated in Fig. 2c, reveal a pronounced two-fold modulation of the magneto-resistance between the superconducting and normal states. In contrast, this two-fold modulation vanishes as the sample transitions into the normal state, indicating a nontrivial origin for this two-fold symmetry[22–25,31]. Additionally, we investigated the angular dependence of the critical magnetic field, and the results consistently displayed a two-fold periodicity at various temperatures as shown in Fig. 2d.

The observation of the two-fold superconducting symmetry puts constraints on the possible paring symmetries of the superconducting order parameters. This is further substantiated by the anisotropic modulation of the superconducting gap[25], which can be revealed through the measurement of the critical current $I_c$. Fig. 3a shows the color mapping of the differential resistance modulated by direction-dependent in-plane magnetic fields. The unambiguous two-fold periodicity of the anisotropic superconducting gap was observed. As the magnetic field increases from 100 mT to 300 mT, the gap diminishes while maintaining its consistent two-fold modulation relative to the magnetic field direction. A waterfall plot of the differential resistance under a magnetic field of 100 mT is depicted in Fig. 3b, where a conspicuous two-fold modulation of the superconducting gap is direct and clear. With an increase in temperature, the superconducting gap decreases, yet the anisotropic nature of the superconducting gap persists until the superconductivity is ultimately suppressed.

Hence, the presence of two-fold rotational symmetry in the superconducting states of $RbV_3Sb_5$ samples is supported by the evidence from magnetotransport experiments, the critical magnetic field measurements, and the observation of an anisotropic superconducting gap. Beyond the local configuration, we extended our investigation to nonlocal configuration measurements by applying a current along terminals 2&6 and simultaneously measuring the voltage between terminals 3&5. This alternative approach revealed an additional six-fold component superimposed on the two-fold symmetry, as illustrated in Fig. 4a. We further show the polar maps of the anisotropic response of magneto-resistivity in Fig. 4b. When we extract the resistance at ~50% of the normal resistance $R_N$, the two-fold symmetry dominates. However, at ~90% $R_N$, a pronounced six-fold component with sharp kinks was observed. Such results suggest that the six-fold components were found near the boundary of the normal states, consistent with the inherent six-fold symmetry of the crystal lattice. However, as the superconducting order becomes dominant, the symmetry was dominant by the two-fold symmetry.

**Ruling out trivial origins**

To eliminate any potential trivial origins, we conducted a series of carefully controlled experiments. We addressed the possibility of out-of-plane field contributions resulting from misalignment between the in-plane field and the sample plane. First, our sample was measured within an Oxford dilution refrigerator equipped with a vector magnet[25]. This setup allowed us to keep the sample fixed while applying only the vector magnetic field, ensuring precise control of the sample angle to a much higher accuracy (better than 0.1°) compared to the traditional mechanical rotation system. Before conducting experiments, we performed system-wide calibration checks to confirm that the magnetic field was applied within the sample plane. Second, we conducted the out-of-plane angle dependence of the critical magnetic field. In Fig. S3, it is clear that the critical magnetic field peaks at $\phi = 90°$, namely the in-plane field, with an accuracy of 0.1°. Third, we considered the scenario in which the in-plane two-fold anisotropy might be influenced by out-of-plane field contributions and calculated the potential angle deviation[32]. Taking the two-fold magnetoresistance at 200 mT (Fig. 2c) as an example, the out-of-plane field dependence, as presented in Fig. 1f, suggests that the out-of-plane field should exceed 20 mT. Consequently, the misalignment angle would need to exceed arcsin(20/200), approximately 5.74°, which is nearly two orders of magnitude larger than our measurement accuracy. Fourth, multiple devices also show similar behavior as presented in Fig. S4.

It has been suggested that the anisotropy may arise from current-induced vortex motion[33]. Such a scenario was ruled out by different configurations as shown in Fig. S5. When the current was applied along $\theta = 90°$ rather than $\theta = 0°$, the angular dependence of the magnetoresistance exhibited the same two-fold symmetry, irrespective of the current direction. Thus, the anisotropy is independent of the current direction. The symmetry evolution in Fig. 4 also rules out the trivial origins such as structural distortion.

**Discussion**

Now we discuss the mechanism behind the two-fold symmetric superconductivity observed in the thin-flake kagome superconductor. Given that charge density waves (CDW) and nematic states have previously been observed in the normal state of the AVS family, the first natural explanation is that such two-fold symmetry in the superconductivity comes from the charge density wave (CDW) scenario. In the three-dimensional CDW configuration, there is a π phase shift between neighboring layers and thus the original six-fold symmetry is lowered to a two-fold one[34–37]. The CDW phase with two-fold symmetry would gap the density of states at the Fermi level, causing the Fermi surface to become truncated with two-fold symmetry. The presence of the six-fold components superimposed on the two-fold symmetry, particularly near the boundary of the normal states (as shown in Fig. 4), further implies the existence of this truncated Fermi surface.

However, it is essential to note that even if two-fold symmetry exists in the normal states, it doesn't automatically guarantee its presence in the superconducting states. For a system with inversion symmetry, as is the case with the AVS family, the conventional spin-orbit coupling terms that only involve spin and momentum degrees of freedom are typically forbidden. Additional parameters, such as parity, are required to induce the anisotropic critical magnetic field[38]. This leads to the expectation of a novel form of unconventional superconductivity known as spin-orbit-parity coupled superconductivity[38,39] in the AVS system. In a manner akin to unconventional superconductivity observed in inversion-symmetric materials like 2M-$WS_2$ (space group $C_{2/m}$) which exhibits a two-fold symmetry[39], the superconducting states in our case could similarly involve parity to account for the observed two-fold symmetric superconducting behavior.

It has been reported the unusual superconducting pairing order exists in AVS[6,17,40]. Given that both nematic instabilities[6,41] and superconducting orders[40,42] have been identified within this family of kagome superconductors, such two orders are also likely to intertwine with each other and give rise to exotic nematic superconductivity, which manifests itself as two-fold symmetry in the transport experiments. Such two-fold symmetry has been observed in doped topological insulators $Cu_xBi_2Se_3$[22,23,26,43–45], iron-based superconductors[20,46–49], twisted graphene[25,50], and transition metal dichalcogenides[38] as evidence of unconventional superconductivity. The gradual evolution of the six-fold symmetric components to two-fold symmetric components in Fig. 4 may further imply additional electronic symmetry breaking or unconventional superconducting pairing when the sample enters into the superconducting states from the normal states. Such a scenario is possible in the presence of a novel multi-component nematic Cooper pairing, which spontaneously breaks the crystalline in-plane rotation symmetry[26,51]. In conjunction with $AV_3Sb_5$'s $D_{6h}$ point group symmetry, four symmetry channels naturally feature two-component superconducting order parameters: $E_{1g}$, $E_{2g}$, $E_{1u}$, and $E_{2u}$, where the first two are even-parity while the last two are odd-parity pairings. The nematic state is formed if only one of the two components in each channel condenses below the superconducting transition, or if the two components condense into a time-reversal invariant linear superposition. For example, the nematic state of the $E_{1u}$ channel would be characterized by either $p_x$ or $p_x+p_y$ pairings and their likes. Notably, based on the existing knowledge about the electronic structure of the $AV_3Sb_5$ family, the nematic state of all these four channels shall generically exhibit nodal gap structure, as would be consistent with recent experiments[52,53]. Therefore, such intertwined orders make the formation of unconventional superconductivity possible in the kagome materials. Other scenarios such as the mixing of s-wave and d-wave/p-wave states[24] or spin-triplet states[19] still need further evidence. Further investigations are still required to determine the full pairing symmetry in the unconventional superconducting phases of RbV$_3$Sb$_5$.

In summary, we systematically observed the two-fold modulation of superconducting states in the kagome superconductor RbV$_3$Sb$_5$. Our results reveal that the kagome superconductor RbV$_3$Sb$_5$ provides a promising platform for unconventional superconductivity, paving the way for fascinating prospects of exploring its interactions with the nontrivial topological properties within this family of materials.

## Methods

### Crystal growth and sample fabrication

Single crystals of $RbV_3Sb_5$ were synthesized via the conventional self-flux method. The thin films of $RbV_3Sb_5$ were exfoliated by polydimethylsiloxane (PDMS) from bulk crystal onto pre-patterned Si substrates with a 285 nm oxide layer. The pre-patterned metal electrodes were prepared by standard electron-beam lithography (EBL) method, followed by deposition of Ti/Au (10 nm/50 nm) using electron-beam evaporation. Finally, the device was encapsulated by hBN to prevent any degradation due to air atmosphere. All device fabrication procedures mentioned above were conducted in the nitrogen-filled glovebox with oxygen and water levels below 0.01 ppm to minimize sample degradation.

### Quantum transport measurements

The samples were measured in an Oxford dilution refrigerator Triton XL1000 with a vector magnet. Thus, the sample can be fixed while a three-dimensional vector magnetic field is applied to realize the angle dependence of the magnetic field. Such a technique greatly improves the angle precision, avoiding the return difference and heat problems of mechanical rotators. The dV/dI measurements were conducted with a rather small alternating current (i.e. 100 nA) while applying a much larger direct current (i.e. 10 µA). A typical lock-in technique and necessary low-temperature filters were used to increase the signal-to-noise ratio of the experiments.

## Data availability

The data that support the plots within this paper and other related findings are available from the corresponding author upon reasonable request.

## Acknowledgments

B.-C. L. thanks Ning Kang, Xiaosong Wu, Yue Zhao, and Kam Tuen Law for valuable discussions. This work was supported by the National Key Research and Development Program of China (Grants No. 2020YFA0309300, No. 2022YFA1403700), the National Natural Science Foundation of China (Grants No. 12074162, No. 12004158, No. 91964201), the Key-Area Research and Development Program of Guangdong Province（Grant No. 2018B030327001), Guangdong Provincial Key Laboratory (Grant No.2019B121203002), Guangdong Basic and Applied Basic Research Foundation (Grant No. 2022B1515130005) and Guangdong province 2020KCXTD001.

## Author contributions

B.-C.L. and S.W. proposed and designed the research. S.W. and B.-C.L. did the quantum transport experiments. S.W., J.-Z.F., and Z.-N.W. fabricated and characterized the sample with the help of Z.W., Y.Y, J.-J.Y., and other authors. S. L and T.C. conducted the EBSD measurements. B.-C.L. and D.Y. supervised the whole project. B.-C.L., S.W., and W.H. wrote the manuscript with inputs from all authors.

## Competing interests

The authors declare that they have no competing interests.

**Figures and Tables**

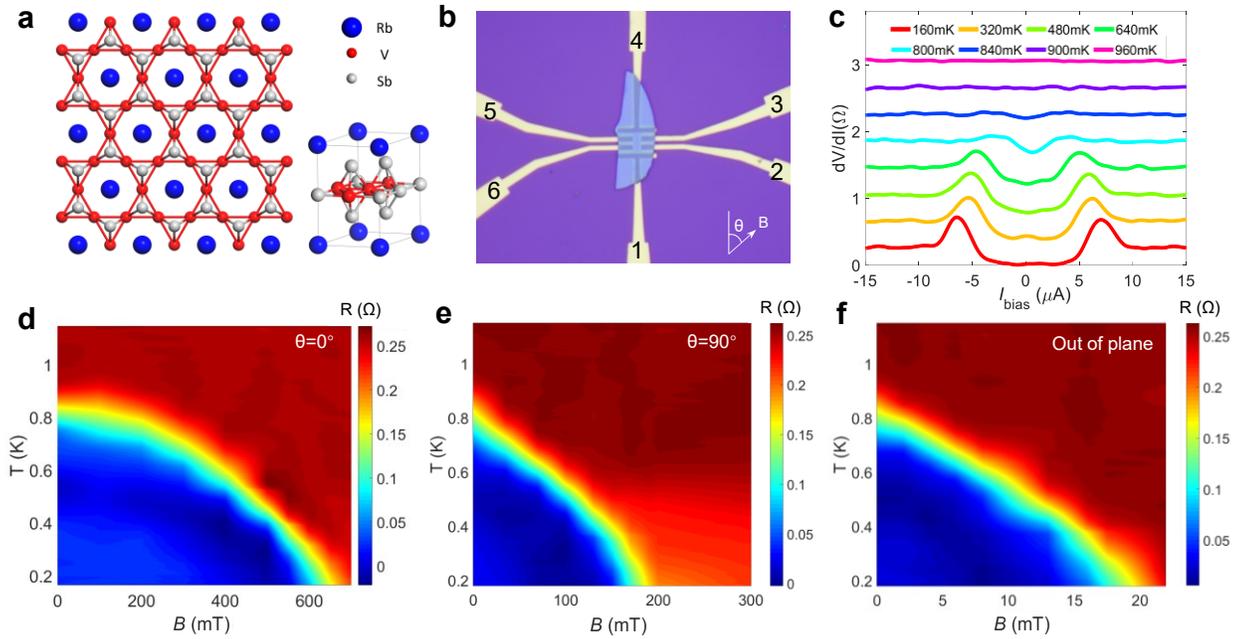

**Fig. 1 The device configuration and superconducting behavior of RbV$_3$Sb$_5$. a** The schematics of the crystal structure for the kagome lattice. **b** The optical image of the RbV$_3$Sb$_5$ device. θ denotes the angle between the magnetic field and the current direction. **c** The differential resistance dV/dI versus bias current I as a function of the temperature shows the typical temperature dependence of the superconducting states. The curves were shifted for clarity. **d-f** The temperature dependence of the magneto-resistance $R_{14,23}$ along different directions. The color bar represents the magnitude of the resistance.

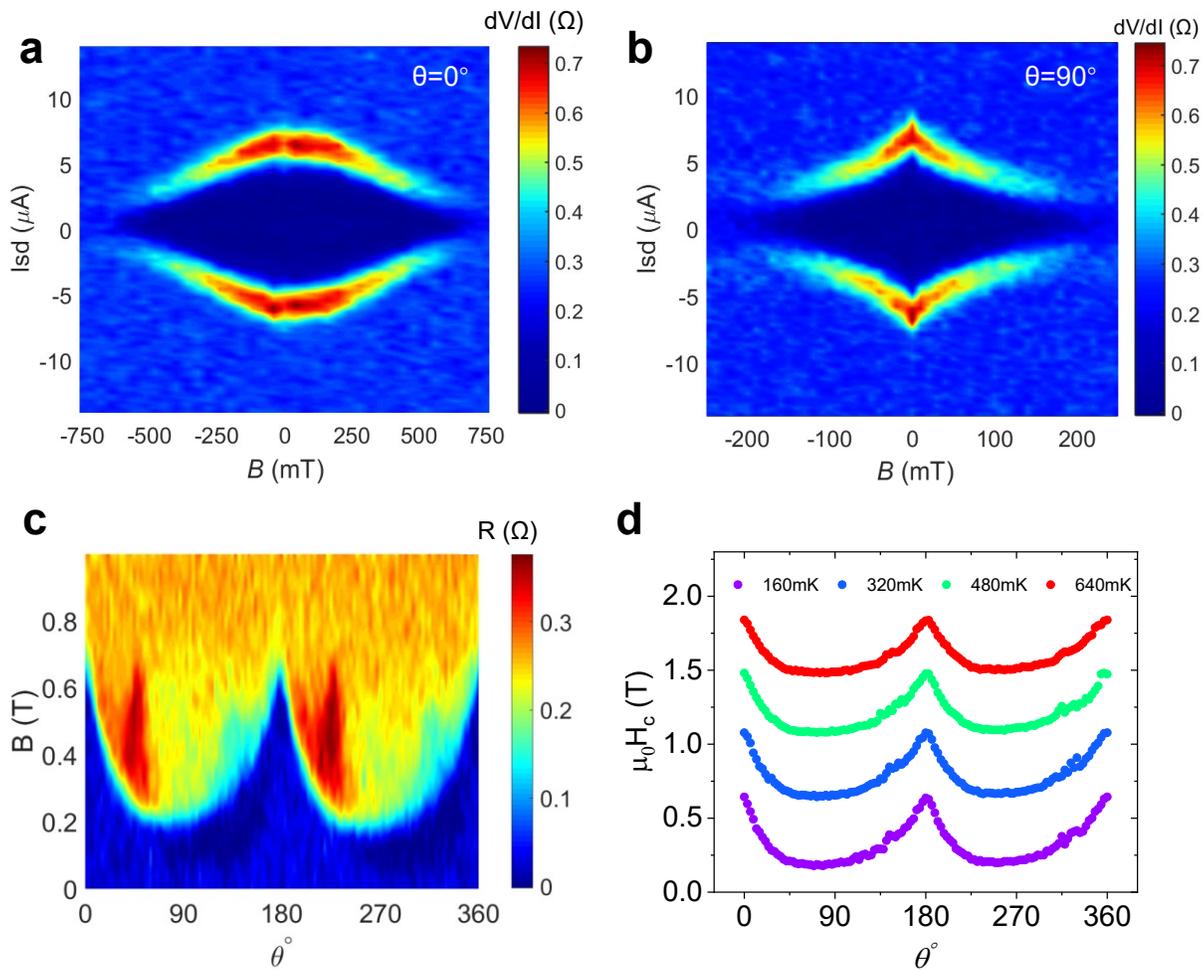

**Fig. 2 The two-fold symmetry of the superconducting states. a,b** Differential resistance versus bias current and magnetic field for θ = 0° and 90°, respectively. The temperature is 160 mK, and the color bar represents the magnitude of the differential resistance. **c** The in-plane angle dependence of the magneto-resistance shows the two-fold symmetry, where θ = 0° denotes that the magnetic field is applied along the current direction and θ = 90° denotes the magnetic field is perpendicular to the current direction. The temperature is 160 mK. **d** The in-plane angle-dependent critical magnetic field $\mu_0 H_c$ at different temperatures also shows the characteristic two-fold anisotropy.

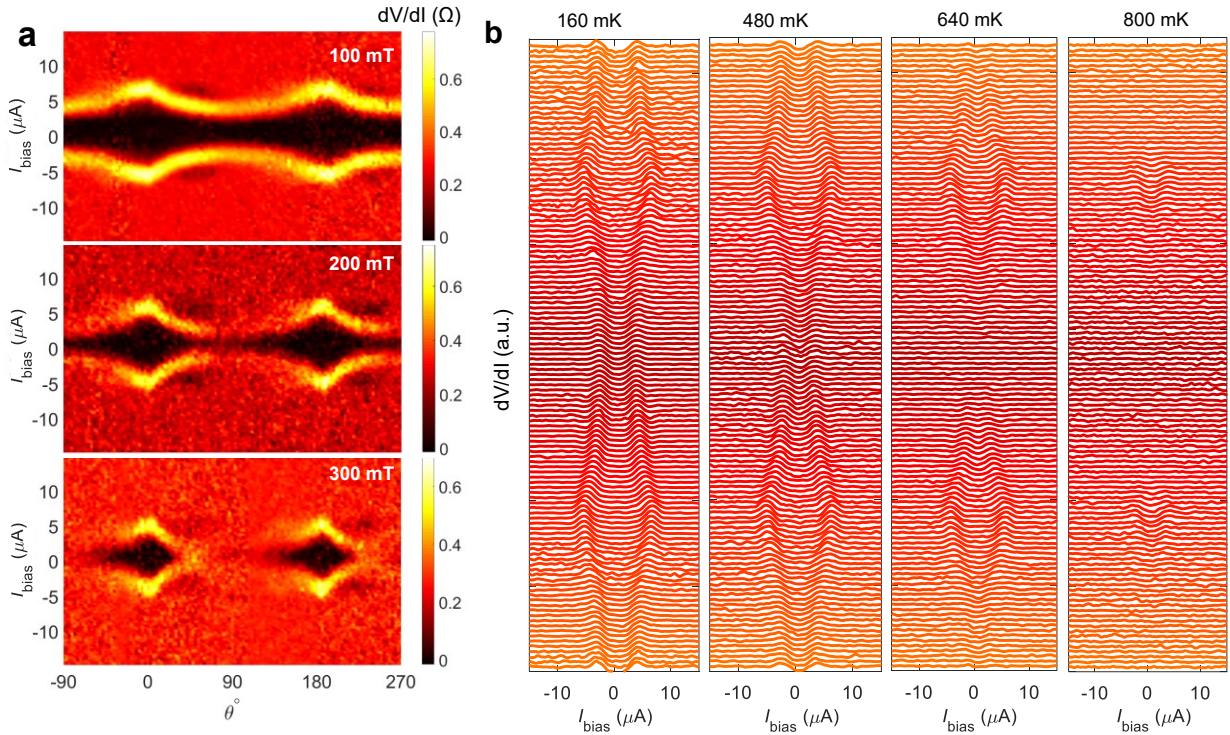

**Fig. 3 The anisotropic response of the superconducting current. a** The differential resistance versus bias current as a function of the direction of the in-plane magnetic field. Increasing the magnetic field strength would gradually close the superconducting gap as expected. The temperature is 160 mK, and the color bar represents the magnitude of the differential resistance. **b** The waterfall plot of the differential resistance at B = 100 mT with angle ranging from -90° to 270°. The superconducting gap diminishes with temperature increasing, while the two-fold symmetry persists as long as the superconducting states exist.

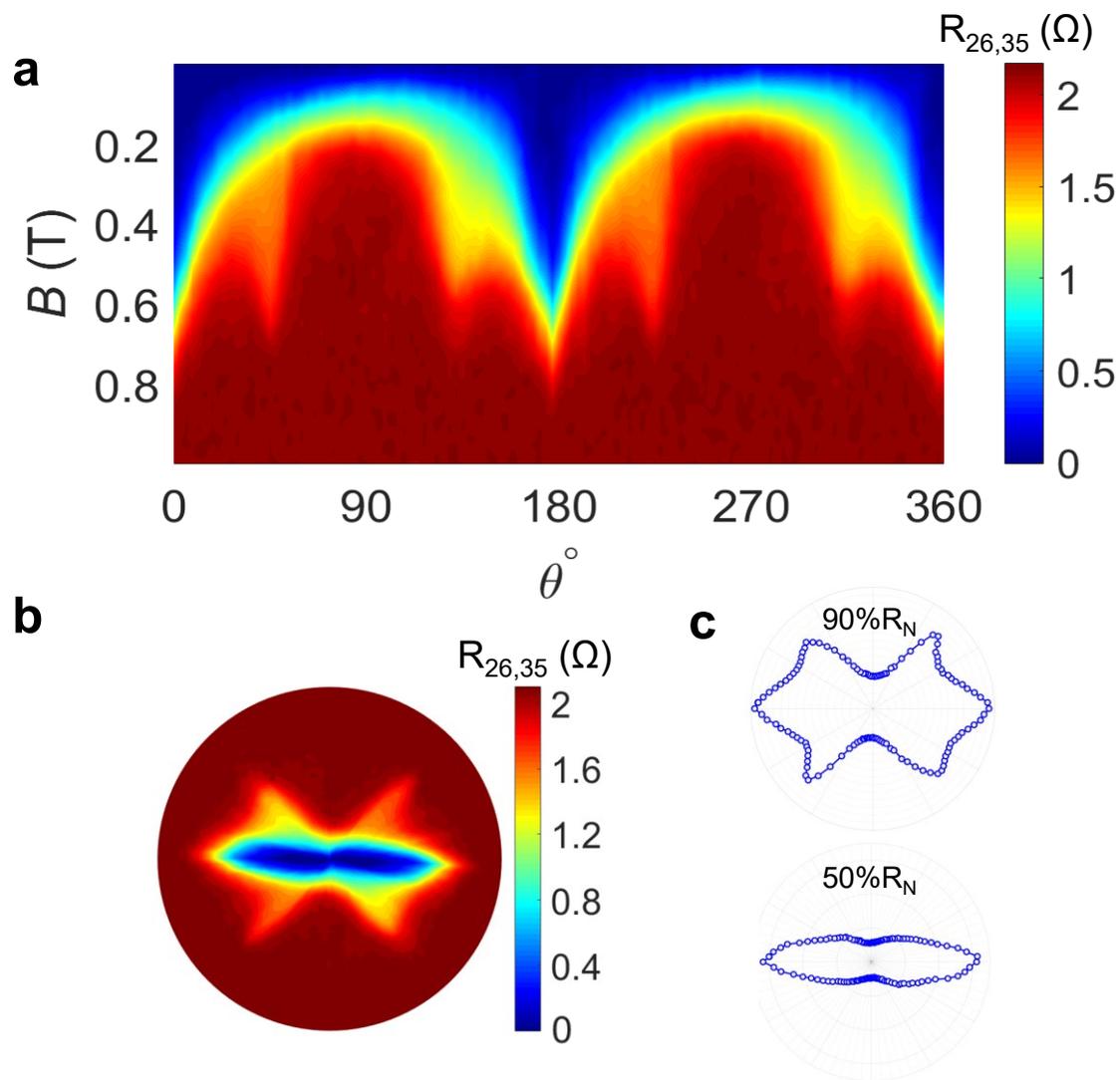

**Fig. 4 The anisotropic magneto-resistance with the nonlocal configuration. a** The in-plane angle dependence of the magneto-resistance shows the six-fold symmetric components superimposed on the two-fold symmetry. The temperature is 160 mK, and the color bar represents the magnitude of the resistance. **b** The polar map of the anisotropic response of the magnetoresistivity. **c** The specific polar plot of the anisotropic magneto-resistance at 90% and 50% of $R_N$, respectively.

# Supplementary Materials for

# Two-fold symmetric superconductivity in the thin-film kagome superconductor $RbV_3Sb_5$

*Corresponding author. Email: linbc@sustech.edu.cn

**This PDF file includes:**

Fig. S1 The temperature dependence of the resistance of a typical $RbV_3Sb_5$ device (S3).

Fig. S2 Differential resistance versus bias current and out-of-plane magnetic fields.

Fig. S3 The out-of-plane angle dependence of the critical magnetic field.

Fig. S4 Another sample S2 showing a similar two-fold symmetric behavior.

Fig. S5 The two-fold superconducting behavior with the current along $\theta = 90°$.

Fig. S6 The in-plane angle dependence of the magneto-resistance of device S1&S2 at a high temperature of 20 K.

Fig. S7 The in-plane angle dependence of the magnetoresistance of device S1 at a high magnetic field of 7.5 T.

Fig. S8 The electron backscatter diffraction (EBSD) pattern of a typical $RbV_3Sb_5$ single crystal.

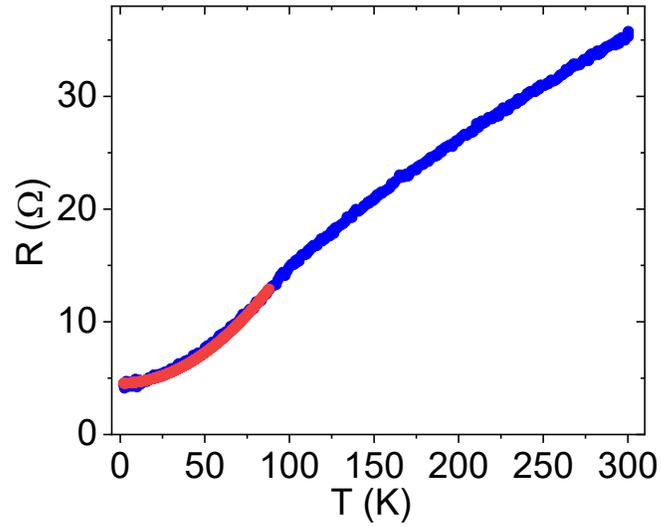

**Fig. S1 The temperature dependence of the resistance of a typical RbV$_3$Sb$_5$ device (S3).** At low temperatures, the resistance can be fitted (red line) according to the Fermi liquid model R=R$_0$+AT$^2$, where T is the temperature, R$_0$ and A are constants, indicating the electron-electron scattering dominates over the electron-phonon scattering. The hump is around 100 K, which is consistent with the critical temperature of the CDW transition.

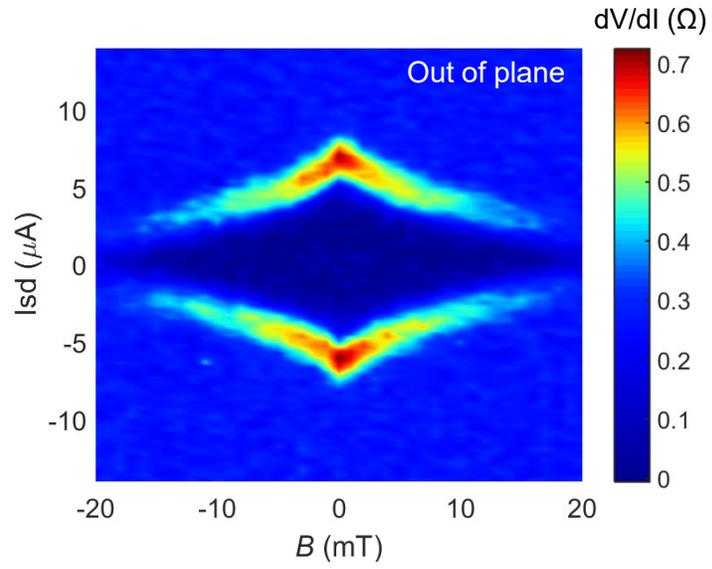

**Fig. S2 Differential resistance versus bias current and out-of-plane magnetic fields.** The out-of-plane anisotropic superconducting behavior is common in thin-film superconductors.

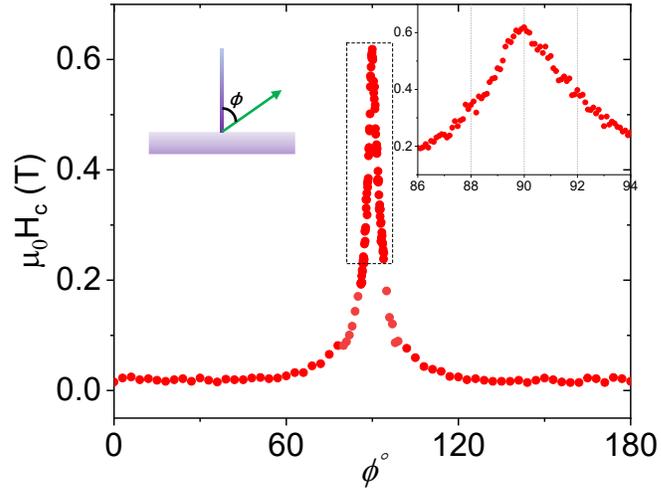

**Fig. S3 The out-of-plane angle dependence of the critical magnetic field.** Here the angle $\phi$ denotes the deviation from the normal direction perpendicular to the sample plane as shown in the inset. The green arrow represents the magnetic field direction. The critical magnetic field peaks at 90°, namely the in-plane direction. Such measurements rule out the possible out-of-plane contribution with an accuracy of ~ 0.1°.

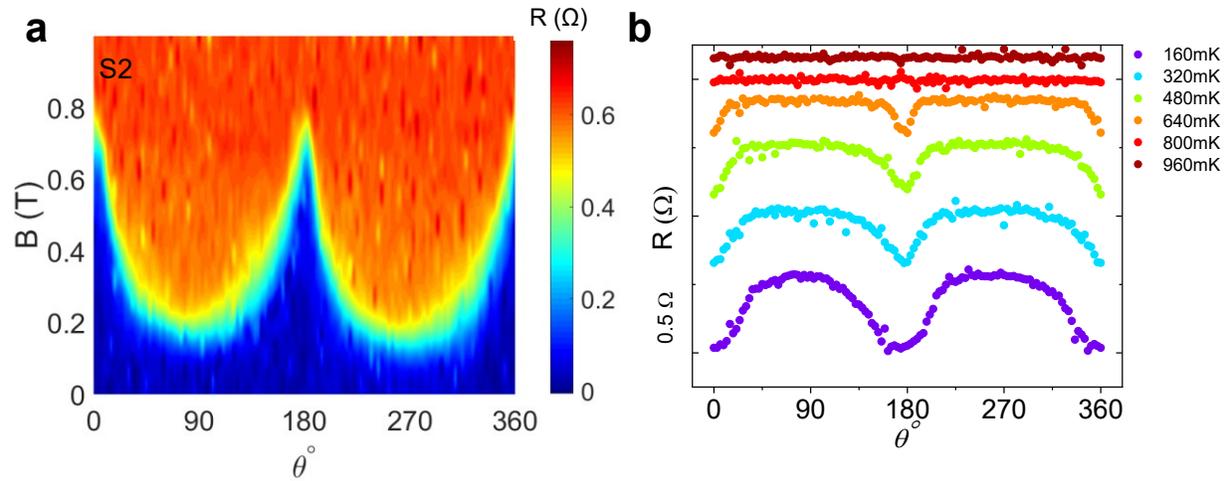

**Fig. S4 Another sample S2 showing a similar two-fold symmetric behavior. a** The in-plane angle dependence of the magneto-resistance was measured. The two-fold anisotropy is only observed when the sample is in the superconducting state. **b** The temperature dependence of the two-fold symmetric magnetoresistance at 300 mT. When the sample is in the normal state at high temperatures, the two-fold symmetry disappears. The curves were shifted for clarity.

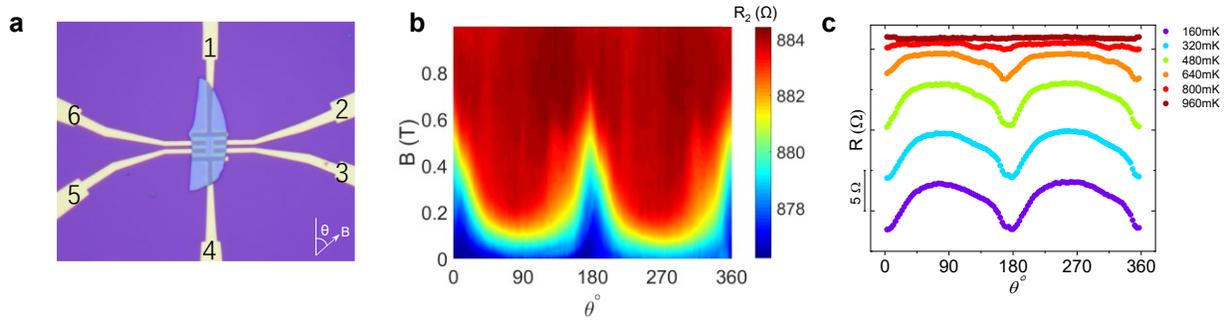

**Fig. S5 The two-fold superconducting behavior with the current along θ = 90°. a** The optical image of device S1 with the number labeling the measurement terminals. **b** The current was applied along terminals 3,5 (perpendicular to the current direction in the main text) and the voltage was simultaneously measured, then we get the two-terminal resistance $R_2$. The two-fold symmetry of the superconducting behavior was also observed, indicating such two-fold symmetry is independent of the current direction. **c** The temperature dependence of the two-fold modulation of the magnetoresistance at 200 mT. The curves were shifted for clarity.

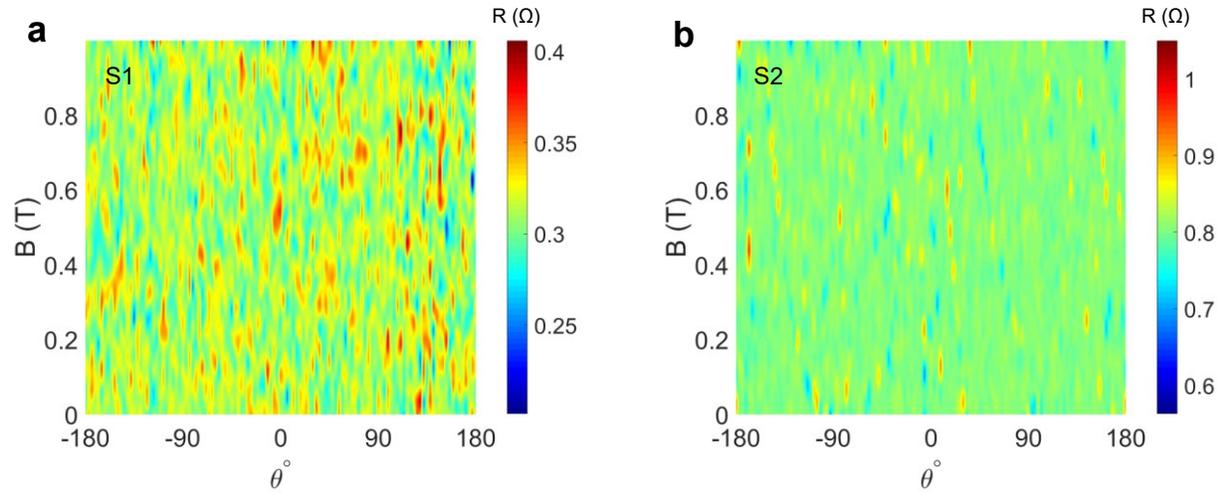

**Fig. S6** The in-plane angle dependence of the magneto-resistance of device S1&S2 at a high temperature of 20 K, far above the critical superconducting temperature. No evident symmetric behavior in any form was observed.

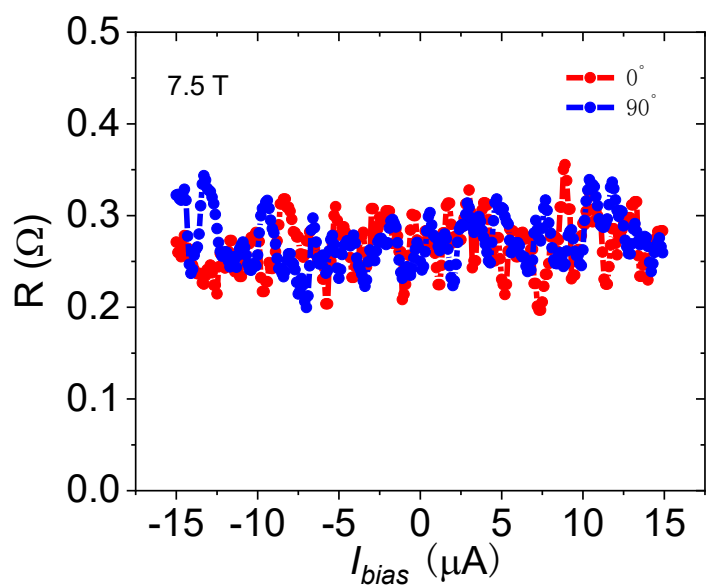

**Fig. S7** The in-plane angle dependence of the magnetoresistance of device S1 at a high magnetic field of 7.5 T, far above the critical magnetic field. No evident anisotropy was observed.

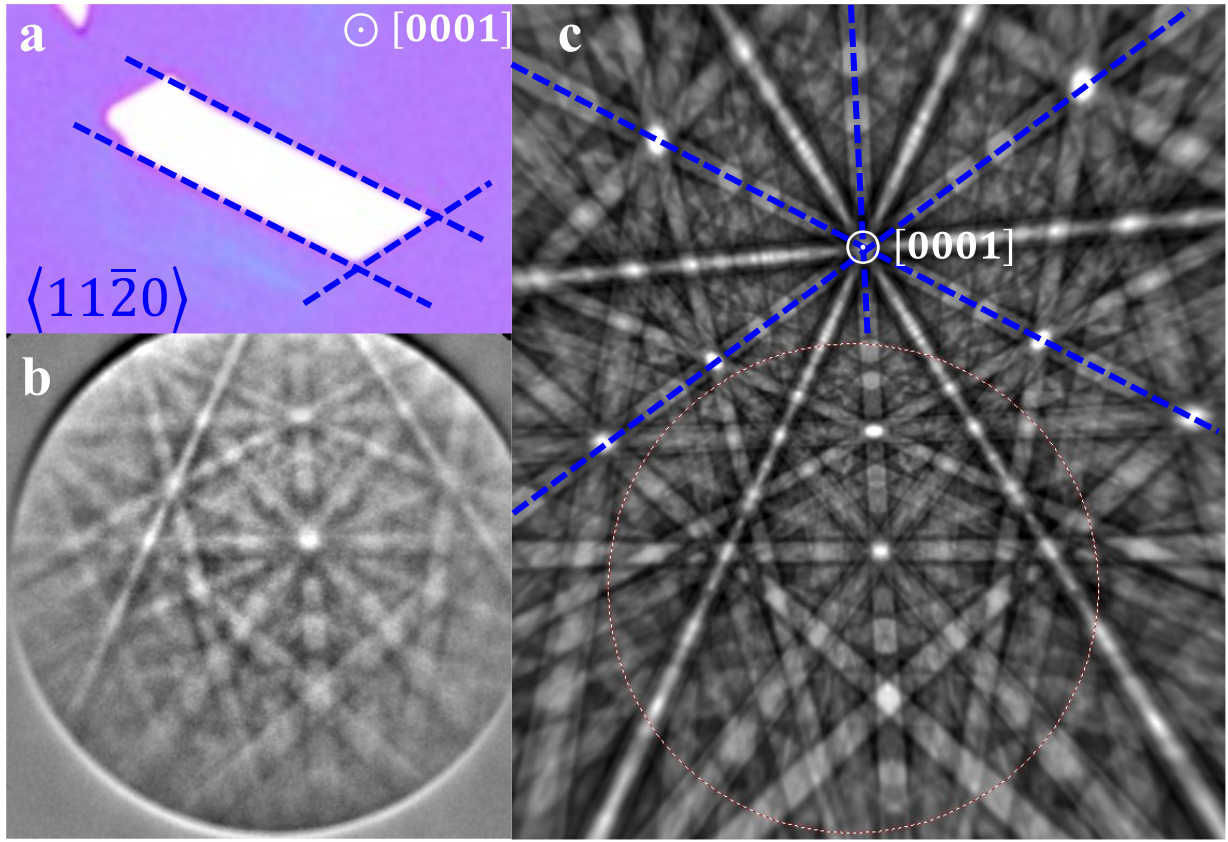

**Fig. S8 The electron backscatter diffraction (EBSD) pattern of a typical RbV$_3$Sb$_5$ single crystal**. **a** The optical image of a typical cleaved sample. **b** The experimental EBSD pattern. **c** The simulated pattern. The simulated pattern in **c** matches the experimental pattern in **b** very well. The blue bands are a family of $\{1\bar{1}00\}$ crystal planes with six-fold rotational symmetry, of which the orientation matches the edges of the RbV$_3$Sb$_5$ flakes shown in **a**. The lateral surfaces of the exfoliated RbV$_3$Sb$_5$ single crystal are therefore the family of $\{1\bar{1}00\}$ crystal planes. The crack propagation direction is $\langle 11\bar{2}0 \rangle$.